%% file: acl_latex.tex
\documentclass[11pt]{article}

\usepackage[final]{acl}

\usepackage{times}
\usepackage{latexsym}

\usepackage[T1]{fontenc}

\usepackage[utf8]{inputenc}

\usepackage{microtype}

\usepackage{inconsolata}

\usepackage{graphicx}

\usepackage{amsmath}
\usepackage{amssymb}
\usepackage{algorithm}
\usepackage{algorithmic}

\usepackage{booktabs}
\usepackage{multirow}
\usepackage{booktabs}
\usepackage{multirow}
\usepackage[table]{xcolor}
\usepackage[normalem]{ulem} 

\definecolor{DarkGreen}{RGB}{220,245,220}
\definecolor{LightGreen}{RGB}{240,252,240}
\definecolor{DarkRed}{RGB}{250,225,225}
\definecolor{LightRed}{RGB}{255,245,245}

\newcommand{\deltaup}[2]{%
  \cellcolor{#1}{\ensuremath{\uparrow}\,#2}%
}

\newcommand{\deltadown}[2]{%
  \cellcolor{#1}{\ensuremath{\downarrow}\,#2}%
}

\usepackage{listings}
\usepackage{xcolor}
\usepackage{mdframed}

\lstdefinelanguage{prompt}{
  basicstyle=\ttfamily\footnotesize,
  breaklines=true,
  frame=single,
  columns=fullflexible,
  keepspaces=true,
}

\title{Improving Retrieval-Augmented Generation without Taxonomy-based Error Categorization}

\author{Gongbo Zhang \\
  Columbia University \\
  \texttt{gz2366@cumc.columbia.edu} \\\And
  Yifan Peng\thanks{Equal contribution corresponding authors.} \\
  Weill Cornell Medicine \\
  \texttt{yip4002@med.cornell.edu} \\\And
  Chunhua Weng\footnotemark[1] \\
  Columbia University \\
  \texttt{cw2384@cumc.columbia.edu} \\
  }

\begin{document}
\maketitle
\begin{abstract}
Retrieval-Augmented Generation (RAG) improves the factual accuracy of large language model (LLM) outputs by grounding generation in external knowledge. Recent agentic RAG systems extend this paradigm with critical agents to evaluate model responses and iteratively refine outputs. However, most prior work implicitly assumes reliable critic feedback and focuses on planning strategies, while paying limited attention to the robustness of the error-correction process itself, which can be impacted by misaligned error categories and ineffective or incorrect corrections. Here, \textit{we hypothesize that RAG performance can be improved without explicit error categorization}. We propose RePAIR, a response–action learning paradigm that directly maps flawed RAG outputs to error-mitigating action plans without relying on fine-grained error taxonomies and explicit critic supervision. Across multiple benchmarks, RePAIR consistently improves agentic RAG performance.
\end{abstract}

\section{Introduction}

Large language models (LLMs) achieve strong generative performance but remain prone to factual hallucinations, which hinder their reliable deployment. Retrieval-Augmented Generation (RAG) partially mitigates this issue by grounding model outputs in external knowledge~\cite{DBLP:conf/nips/LewisPPPKGKLYR020, DBLP:journals/jmlr/IzacardLLHPSDJRG23, DBLP:journals/tmlr/MialonDLNPRRSDC23, DBLP:conf/kdd/FanDNWLYCL24, zhang2025longcontextrag, DBLP:conf/acl/XuMDZW025}. More recently, agentic RAG systems extend this paradigm by introducing critic agents that evaluate model outputs and guide iterative actions over retrieval and generation~\cite{DBLP:conf/iclr/AsaiWWSH24, DBLP:conf/iclr/0002WSLCNCZ23, DBLP:conf/nips/YaoYZS00N23, DBLP:journals/corr/abs-2401-15884, DBLP:conf/acl/DongJL0DW25, fang2026generativeqe}. In this work, we adopt the narrower definition of agentic RAG used in prior refinement-based systems, focusing on iterative corrective planning over structured retrieval and generation operations rather than open-ended tool use or web-scale environments.

Although critic-guided refinement can improve end-to-end performance, most approaches assume the critic is reliable and focus on \emph{using} critic feedback for planning modules ~\cite{DBLP:conf/nips/MadaanTGHGW0DPY23, DBLP:conf/nips/YaoYZS00N23, DBLP:conf/nips/ShinnCGNY23}, error taxonomies~\cite{DBLP:conf/acl/DongJL0DW25}, and multi-step control flows~\cite{DBLP:conf/iclr/AsaiWWSH24, DBLP:conf/www/0002LJND24, DBLP:journals/corr/abs-2401-15884, DBLP:conf/emnlp/KimL24}. In practice, LLM-based critics frequently misidentify failure causes and recommend ineffective corrections that can sometimes degrade performance relative to non-agentic baselines~\cite{DBLP:conf/nips/ZhengC00WZL0LXZ23, DBLP:conf/acl/WangLCCZLCKLLS24, DBLP:conf/iclr/0009CMZYSZ24, DBLP:conf/emnlp/LiuIXWXZ23, DBLP:conf/emnlp/KimL24}. These observations suggest that reliance on explicit error categorization may introduce additional sources of uncertainty under noisy retrieval and generation.

This raises a key question: \emph{is explicit error categorization essential for effective RAG improvement}? Existing agentic RAG systems typically treat critic-generated error categories as a central intermediate representation for planning and correction~\cite{DBLP:conf/iclr/AsaiWWSH24, DBLP:conf/acl/DongJL0DW25, DBLP:journals/corr/abs-2401-15884, DBLP:conf/nips/YaoYZS00N23, DBLP:conf/nips/ShinnCGNY23}. We hypothesize that \textit{explicit error categorization is not essential to improve RAG performance}.

To test this, we introduce \textbf{RePAIR}, a RAG Response-Action learning paradigm that directly maps flawed RAG outputs to effective actions without relying on detailed error taxonomies or explicit critic supervision. RePAIR treats error categorization as a latent process and learns a policy over actions conditioned on the RAG state, thereby unifying error categorization and planning into a single learning objective. Our goal is not to invalidate taxonomy-based approaches, but to demonstrate that competitive corrective performance can be achieved without requiring explicit intermediate error categorization. 

Experiments on three benchmark datasets show that RePAIR improves token-level F1 by 3.8 points (13.1\%) over standard RAG and outperforms all agentic RAG baselines.

Our contributions are three-fold:
\begin{itemize}
    \item We propose a response–action learning formulation for RAG refinement that removes the need for explicit error categorization and unifies diagnosis and planning within a single policy.
    \item We introduce a two-phase training strategy that combines oracle-guided off-policy bootstrapping with on-policy refinement under deployment-matched conditions.
    \item We empirically show that competitive and stable performance can be achieved without explicit error taxonomies across multiple QA benchmarks.
\end{itemize}

\section{Related Work}

Our work bridges two directions: agentic RAG and evaluation-driven planning and reasoning in LLMs.

RAG improves factual reliability by grounding language model outputs in external knowledge~\cite{DBLP:conf/nips/LewisPPPKGKLYR020, DBLP:journals/jmlr/IzacardLLHPSDJRG23, DBLP:journals/tmlr/MialonDLNPRRSDC23, DBLP:conf/kdd/FanDNWLYCL24}. Recent \textbf{Agentic RAG} explicitly categorizes errors and implements mitigating mechanisms. For example, \citet{DBLP:conf/iclr/AsaiWWSH24} integrated self-reflection via control tokens that regulate retrieval, relevance assessment, and filtering. Similarly, \citet{DBLP:conf/www/0002LJND24} separated cognition from metacognitive regulation by diagnosing knowledge-related failures and planning targeted corrections. \citet{DBLP:journals/corr/abs-2401-15884} employed an external evaluator to assess evidence reliability and trigger additional retrieval under noisy conditions. Across these approaches, critic-generated evaluations act as intermediate representations that guide subsequent retrieval or regeneration. \citet{DBLP:conf/acl/DongJL0DW25} formalized this design using a hierarchical error taxonomy combined with critic-guided action planning, while \citet{DBLP:conf/nips/RuQHZSCJWSLZWJ024} provided fine-grained diagnostic evaluations without learning policies. Collectively, these methods exemplify error categorization-based agentic RAG pipelines.

LLMs can also use evaluation signals to guide multi-step \textbf{planning and reasoning}. Prior work has framed reasoning as a search over intermediate states controlled by LLM-based evaluators \cite{DBLP:conf/nips/YaoYZS00N23}, stored verbal self-critiques to inform subsequent actions \citet{DBLP:conf/nips/ShinnCGNY23}, or applied iterative self-feedback to improve outputs \cite{DBLP:conf/nips/MadaanTGHGW0DPY23}. In parallel, reinforcement learning approaches learn action policies from outcome-based signals for long-horizon planning~\cite{DBLP:journals/corr/abs-2503-14476, DBLP:journals/corr/abs-2510-05592}. Tool-augmented reasoning or multi-agent frameworks \cite{DBLP:conf/iclr/HongZCZCWZWYLZR24, DBLP:journals/corr/abs-2308-08155} emphasize general agent capabilities or structured coordination. 

Although effective, these approaches rely on explicit error categorization or multi-step reasoning and assume reliable self-evaluation, which can lead to instability when feedback is unreliable. By contrast, our work directly learns response-to-action policies. This eliminates the need for predefined error taxonomies or external evaluators while still capturing latent error signals. Consequently, our method unifies error categorization and planning in a single, learnable framework, enabling more flexible and robust model behavior.

\section{Methodology}

We formulate RePAIR as a two-phase, preference-based policy learning problem. Given an initially failed RAG instance, the goal is to learn a policy that outputs a sequence of high-level RAG operations (i.e., a \emph{plan}) whose execution produces a corrected answer. We distinguish an \emph{off-policy} phase, where explicit correctness and error signals are available to bootstrap learning, from an \emph{on-policy} phase, where the planner must infer failures solely from the raw RAG context.

\subsection{RAG State and Plan Representation}

Let $q$ denote a user question, $D=\{d_1,\dots,d_k\}$ the set of documents retrieved by a baseline retriever, and $a_0$ the initial answer produced by a RAG system. For training instances, we assume access to a ground-truth answer $a_{\text{gold}}$.

During off-policy training, the planner observes an oracle-derived binary correctness label $c\in\{0,1\}$, obtained by comparing $a_0$ with $a_{\text{gold}}$, indicating whether the initial RAG answer is correct. When $a_0$ is incorrect, the planner additionally observes an incorrect reasoning trace $r_0$ produced by the baseline system. The planner, therefore, conditions on the augmented state $x_{\text{off}}=(q,D,a_0,c,r_0)$ during off-policy learning.

During on-policy training, we simulate the inference-time setting where gold answers are unavailable. The planner receives a coarse correctness estimate $\hat{c}\in\{0,1\}$, inferred solely from the RAG response via an LLM-based judge, without access to $a_{\text{gold}}$. The planner thus conditions on the reduced state $x_{\text{on}}=(q,D,a_0,\hat{c})$ during on-policy learning. At inference time, the planner observes only the same reduced state and has no access to any explicit diagnostic or oracle signals.

Let $\mathcal{O}$ denote a predefined set of high-level RAG operations, including \textsc{Rewrite}, \textsc{Decompose}, \textsc{Retrieval}, \textsc{RefineDoc}, and \textsc{GenerateAnswer}. A \emph{plan} is a variable-length sequence of operations $y=(o_1,\dots,o_T)$, where each $o_t\in\mathcal{O}$ and $T$ denotes the length of the plan. Our experiments focus on this structured action space to isolate the effect of removing explicit error categorization; extending the approach to more open and complex action spaces (e.g., tool use or API interaction) is left for future work.

\subsection{Response-Action Plan Policy}

Given a RAG instance $x$ and a plan $y$, an executor deterministically applies the sequence of operations specified by $y$ to produce a revised answer,
\(
    \hat{a}(y;x)=\text{Executor}(x,y)
\).
The revised answer is evaluated against the ground-truth answer $a_{\text{gold}}$ using a scalar reward function $R(x,y)$. Concretely, $R(x,y)\in \mathbb{R}$ is defined as the token-level F1 score between $\hat{a}(y;x)$ and $a_{\text{gold}}$, as described in the experimental setup. The reward thus quantifies the effectiveness of the plan in correcting the initial RAG failure.

For each input $x$, we execute a set of candidate plans $\{y_1,\dots,y_n\}$ and induce pairwise preferences by comparing their rewards, such that \(y_i \succ y_j \), iff $R(x, y_i) > R(x, y_j)$, yielding preference triples $(x,y^{+},y^{-})$, where $y^{+}$ denotes the plan with the higher F1 score.

We learn a conditional plan policy $\pi_\theta(y| x)$ that maps a RAG state to a distribution over plans, and define a reference policy $\pi_{\text{ref}}$ as the initial pretrained model. The planner is trained using Direct Preference Optimization (DPO)~\cite{DBLP:conf/nips/RafailovSMMEF23}. For each preference triple $(x,y^{+},y^{-})$, the training objective is
\begin{equation}
\label{eq:dpo}
\mathcal{L}_{\text{DPO}}
=
-\log\sigma\!\left(
\beta \log
\frac{\pi_\theta(y^{+}| x)/\pi_{\text{ref}}(y^{+}| x)}
     {\pi_\theta(y^{-}| x)/\pi_{\text{ref}}(y^{-}| x)}
\right),
\end{equation}
where $\sigma$ denotes the logistic function and $\beta$ controls the strength of
regularization toward the reference policy.

\paragraph{Preference construction and noise.}
We construct DPO preference pairs using relative reward rankings over candidate plans. In some cases, both plans may achieve low absolute rewards, which could introduce noisy supervision. However, our objective is to optimize relative preference: DPO encourages the model to favor plans that yield comparatively better outcomes under the same conditions. In practice, two factors mitigate the impact of such noise. First, if the learned policy fails to identify a superior plan, the system defaults to the vanilla RAG response, which serves as a lower bound. Second, even when absolute rewards are low, relative differences often reflect variations in grounding or reasoning quality rather than reinforcing systematic hallucinations. Nevertheless, filtering near-tied or uniformly low-reward pairs may further reduce noise and improve robustness, which is beyond the scope of the current work.

\subsection{Two-Phase Training}

We adopt a two-phase training strategy to address a trade-off between learning stability and deployment alignment. Learning corrective plans directly from deployment-style inputs can be unstable, as the reward signal provides only coarse feedback on the entire action sequence rather than step-wise supervision. The \textit{off-policy} phase mitigates this issue by leveraging oracle-derived correctness signals and reasoning traces to bootstrap stable response–action learning. However, reliance on oracle signals introduces a train–test mismatch, since such information is unavailable at inference time. The \textit{on-policy} phase removes these oracle signals and refines the planner under deployment-matched inputs, using only coarse correctness estimates derived from the RAG response. This phase improves robustness by aligning training conditions with inference (Appendix~\ref{app:algorithms}).

\section{Experimental Setup}

We follow the experimental protocol of RAG-Critic~\cite{DBLP:conf/acl/DongJL0DW25} and evaluate RePAIR on three question answering benchmarks (Table~\ref{tab:dataset_sizes}): \emph{Natural Questions} (NQ)~\cite{DBLP:journals/tacl/KwiatkowskiPRCP19}, \emph{Wizard of Wikipedia} (WoW)~\cite{DBLP:conf/iclr/DinanRSFAW19}, and \emph{2WikiMultiHopQA} (2Wiki)~\cite{DBLP:conf/coling/HoNSA20}, using the same evaluation splits. These datasets cover single-hop factual QA, knowledge-grounded dialogue, and multi-hop reasoning, respectively.
{\input{latex/tables/table-dataset}

We compare RePAIR against a range of standard and critic-based RAG baselines, including vanilla RAG without refinement, Self-Refine~\cite{DBLP:conf/nips/MadaanTGHGW0DPY23}, FLARE~\cite{DBLP:conf/emnlp/JiangXGSLDYCN23}, Self-RAG~\cite{DBLP:conf/iclr/AsaiWWSH24}, MetaRAG~\cite{DBLP:conf/www/0002LJND24}, and RAG-Critic~\cite{DBLP:conf/acl/DongJL0DW25}. For fair and controlled comparisons, we adopt Qwen2.5-7B-Instruct~\cite{DBLP:journals/corr/abs-2407-10671} and Llama3.1-8B~\cite{meta-llama-3-1} as a shared backbone across all methods, following the RAG-Critic configuration. Detailed experimental settings and configurations are provided in Appendices \ref{app:details} and \ref{app:dpo}.

We report token-level F1 on all datasets, computed as the lexical overlap between the predicted answer and the set of gold answers after normalization. For each example, the maximum F1 score across gold answers is used.

\section{Results and Discussion}

\subsection{Main Results}

Across all three benchmarks, RePAIR achieves the strongest overall performance (Table~\ref{tab:main_results}). It improves the average F1 by 3.8 points over the vanilla RAG baseline and outperforms all critic-centric agentic RAG methods. In contrast, prior approaches (e.g., Self-Refine, FLARE, and Self-RAG) show mixed or negative gains, sometimes underperforming vanilla RAG. RePAIR demonstrates stable improvements across datasets, achieving the best results on NQ and 2Wiki and the second-best on WoW, indicating robust effectiveness across both single-hop and multi-hop question answering. Compared to RAG-Critic, RePAIR also delivers consistently larger improvements without relying on explicit error taxonomies or critic supervision.

\begin{table}[t]
\centering
\small
\setlength{\tabcolsep}{4pt}
\begin{tabular}{cl@{~~}rrrc}
\toprule
\textbf{\#} & \textbf{Method} & \textbf{NQ} & \textbf{WoW} &
\textbf{2Wiki} & \textbf{Avg. ($\Delta$)} \\
\midrule
1 & \textit{Standard RAG} & 38.3 & 10.2 & 30.1 & 26.2 \\
\midrule
\multicolumn{6}{l}{\textit{Agentic RAG}} \\
2 & Self-Refine  & 22.3 & 11.7 & 23.1 & 19.0 ~~(\deltadown{DarkRed}{7.2}) \\
3 & FLARE        & 19.8 &  4.2 & 22.7 & 15.6 (\deltadown{DarkRed}{10.6}) \\
4 & Self-RAG     & 32.3 & \textbf{17.4} & 18.9 &
22.9 ~~(\deltadown{LightRed}{3.3}) \\
5 & MetaRAG      & 40.2 &  6.2 & 29.2 &
25.2 ~~(\deltadown{LightRed}{1.0}) \\
6 & RAG-Critic   & \textbf{42.0} & 11.6 & 33.1 &
\underline{28.9} ~~(\deltaup{LightGreen}{2.7}) \\
\midrule
\multicolumn{6}{l}{\textit{Ours}} \\
7 & RePAIR (offline) & 38.4 & 14.5 & 31.5 &
28.1 ~~(\deltaup{LightGreen}{1.9}) \\
8 & RePAIR (online)  & 38.8 & 14.5 & \underline{33.0} &
28.8 ~~(\deltaup{LightGreen}{2.6}) \\
9 & RePAIR           & \underline{40.3} & \underline{15.3} &
\textbf{34.5} & \textbf{30.0} ~~(\deltaup{DarkGreen}{3.8}) \\
\bottomrule
\end{tabular}
\vspace{0.4em}
\caption{Comparison between RePAIR and existing agentic RAG frameworks.
Best token-level F1 scores are in \textbf{bold}, and second-best
are \underline{underlined}.}
\label{tab:main_results}
\end{table}

\subsection{Impact of the Two-Phase Training}

We examine the effectiveness of the two-phase training strategy by comparing three variants: off-policy only, on-policy only, and the full model (rows 7--9 in Table \ref{tab:main_results}). Off-policy training alone underperforms because the planner relies on oracle correctness signals that are unavailable at inference time, resulting in a train–test mismatch. In contrast, on-policy training alone operates on deployment-style inputs but lacks sufficient supervision, making it difficult to learn effective corrective policies from sparse and noisy reward signals. The full two-phase RePAIR resolves this trade-off by first stabilizing learning with oracle-guided supervision and then refining the policy under inference-matched conditions. This combination yields better overall performance than either phase alone, indicating that both stages play a key role in effective training. These results suggest that LLM-based critics remain useful when providing coarse correctness signals, but fine-grained error categorization may introduce additional instability.

\begin{table}[t]
\centering
\small
\begin{tabular}{llrrr}
\toprule
\textbf{Dataset} & \textbf{Action} &
\textbf{w/ Err.} &
\textbf{w/o Err.} &
\textbf{$\Delta$ (\%)} \\
\midrule
{\textbf{NQ}}
& Retrieval       &  310 & 138 & -55.5 \\
& Rewrite    &   12 &   0 & -100.0 \\
& Decompose  &    2 &   1 &  -50.0 \\
& RefineDoc       &    3 &   0 & -100.0 \\
\midrule
{\textbf{2Wiki}}
& Retrieval       & 1563 & 491 &  -68.6 \\
& Rewrite    &   86 &  43 &  -50.0 \\
& Decompose  &   14 &  11 &  -21.4 \\
& RefineDoc       &   20 &   7 &  -65.0 \\
\midrule
{\textbf{WoW}}
& Retrieval       &    8 &   7 &  -12.5 \\
& Rewrite    &    0 &   0 &   -- \\
& Decompose  &    0 &   0 &   -- \\
& RefineDoc       &    0 &   0 &   -- \\
\bottomrule
\end{tabular}
\vspace{0.4em}
\caption{Comparison of action usage with and without error categorization optimization across datasets.}
\label{tab:action_analysis}
\end{table}

\subsection{Analysis of Planner Action Usage}

We analyze the planner’s action usage before and after optimization, excluding the final answer-generation step (Table~\ref{tab:action_analysis}). Across all datasets, optimization without explicit error categorization leads to substantially more concise action sequences, with large reductions in retrieval, query rewriting, decomposition, and document refinement. Notably, these reductions do not compromise performance; RePAIR maintains comparable or improved QA results, indicating that effective behavior can be learned without frequent auxiliary actions. Importantly, actions such as query rewriting or document refinement remain valuable but are more effective when applied selectively rather than being triggered by potentially unreliable fine-grained error analysis. Overall, the optimized planner learns a more efficient policy that avoids redundant interventions while still correcting errors.

\subsection{Case Analysis}

Figure~\ref{fig:error_vs_plan} illustrates how error categorization affects planning in agentic RAG. For a factual question on banking regulation in India, the retriever returns passages that cover multiple related statutes, while the initial RAG response fails to identify the correct governing act. With accurate error labels, this failure is recognized as a generator issue, prompting direct regeneration of the answer. In contrast, an incorrect error categorization misattributes the issue to the retriever, triggering unnecessary query rewriting and additional retrieval before regenerating the answer. By avoiding fine-grained misclassification, RePAIR directly selects the appropriate action, yielding more concise and effective plans.

\input{latex/figure-example}

\section{Conclusion}
We presented RePAIR, a response-action learning paradigm for agentic RAG that
eliminates the need for explicit error categorization. By directly optimizing action policies from response–action preferences, RePAIR reduces
dependence on brittle critic judgments and yields more stable refinement under noisy retrieval and generation conditions. Empirical results across multiple benchmarks demonstrate that RePAIR consistently outperforms critic-centric approaches in both accuracy and efficiency.

\section*{Limitations}
While RePAIR demonstrates consistent gains over critic-centric agentic RAG baselines, our study has several limitations. First, experiments are restricted to three open-domain QA benchmarks and a fixed set of high-level RAG operations; performance and stability in other domains or with richer action spaces remain to be explored. Additionally, although RePAIR shows improved stability compared to critic-driven approaches, we do not comprehensively characterize failure modes under noisier and more adversarial retrieval conditions. Addressing these limitations is an important step toward understanding the generality of response–action learning in agentic RAG systems.

\section*{Acknowledgements}
This work was supported by the National Library of Medicine [grant numbers R01LM014344, R01LM014573].

\bibliography{custom}

\clearpage
\newpage

\appendix
\label{sec:appendix}

\section{Detailed RePAIR Algorithm Design}
\label{app:algorithms}

We adopt a two-phase training strategy that leverages diagnostic supervision when available, while avoiding reliance on such signals at inference time. The \textit{off-policy} phase uses explicit correctness labels and reasoning traces to bootstrap stable response–action learning, whereas the \textit{on-policy} phase removes these signals to align training with deployment conditions and ensure robustness without explicit error categorization.

In the \textbf{off-policy phase} (Algorithm~\ref{alg:rag_dpo_off}), we assume access to a static log of RAG instances with correctness labels and reasoning traces:
\begin{equation}
    \mathcal{D}_{\text{off}} = \{(x_{\text{off}}^{(i)}, a_{\text{gold}}^{(i)})\}_{i=1}^{N}.
\end{equation}
For each $x_{\text{off}}^{(i)}$, a teacher model proposes candidate plans $\{y^{(i)}_1, \dots, y^{(i)}_{n_i}\}$ conditioned on the augmented state $(q^{(i)}, D^{(i)}, a_0^{(i)}, c^{(i)}, r_0^{(i)})$. Each plan is executed to obtain a revised answer $\hat{a}(y^{(i)}_j; x_{\text{off}}^{(i)})$ and evaluated using the reward $R(x_{\text{off}}^{(i)}, y^{(i)}_j)$. These scores are used to construct preference triples $(x_{\text{off}}^{(i)}, y^{(i,+)}, y^{(i,-)})$ and minimized using the DPO loss in Eq.~\eqref{eq:dpo}, yielding an off-policy optimized planner $\pi_{\theta}^{\text{off}}$.

In the \textbf{on-policy phase} (Algorithm~\ref{alg:rag_dpo_on}), we further refine the planner \emph{without} providing access to reasoning traces $r_0$. We consider a dataset
\begin{equation}
    \mathcal{D}_{\text{on}} = \{(x_{\text{on}}^{(i)}, a_{\text{gold}}^{(i)})\}_{i=1}^{M},
\end{equation}
where $x_{\text{on}}^{(i)} = (q^{(i)}, D^{(i)}, a_0^{(i)}, \hat{c}^{(i)})$. For each $x_{\text{on}}^{(i)}$, the current planner $\pi_{\theta}$ (initialized from $\pi_{\theta}^{\text{off}}$) generates one or more candidate plans $\{y^{(i)}_1, \dots, y^{(i)}_{k_i}\}$. These plans are executed and evaluated using the same reward function $R(x_{\text{on}}^{(i)}, y^{(i)}_j)$, enabling preference-based optimization under deployment-matched conditions.

\begin{algorithm}
\caption{Off-Policy DPO Training for RAG Planning}
\label{alg:rag_dpo_off}
\begin{algorithmic}[1]
\REQUIRE Off-policy dataset $\mathcal{D}_{\text{off}}$, executor, scoring function $R$, reference policy $\pi_{\text{ref}}$
\STATE $\pi_{\theta} \leftarrow \pi_{\text{ref}}$
\STATE $\mathcal{P}_{\text{off}} \leftarrow \emptyset$
\FOR{each $(x_{\text{off}}, a_{\text{gold}}) \in \mathcal{D}_{\text{off}}$}
    \STATE $x_{\text{off}} = (q, D, a_0, c, r_0)$
    \STATE Use a teacher model to propose candidate plans $\mathcal{Y}$ conditioned on $x_{\text{off}}$
    \FOR{each $y_j \in \mathcal{Y}$}
        \STATE $\hat{a}_j \leftarrow \text{Executor}(x_{\text{off}}, y_j)$
        \STATE $s_j \leftarrow R(x_{\text{off}}, y_j)$
    \ENDFOR
    \STATE Induce preferences over $\mathcal{Y}$ using scores $\mathcal{S}$ (e.g., by ranking) and construct one or more triples $(x_{\text{off}}, y^{+}, y^{-})$
    \STATE Add all resulting triples to $\mathcal{P}_{\text{off}}$
\ENDFOR
\STATE Train $\pi_{\theta}$ on $\mathcal{P}_{\text{off}}$ using the DPO loss in Eq.~\eqref{eq:dpo}
\STATE Denote the resulting planner as $\pi_{\theta}^{\text{off}}$
\STATE \textbf{return} $\pi_{\theta}^{\text{off}}$
\end{algorithmic}
\end{algorithm}

\begin{algorithm}
\caption{On-Policy DPO Refinement for RAG Planning}
\label{alg:rag_dpo_on}
\begin{algorithmic}[1]
\REQUIRE On-policy dataset $\mathcal{D}_{\text{on}}$, executor, scoring function $R$, reference policy $\pi_{\text{ref}}$, initialized planner $\pi_{\theta}^{\text{off}}$, number of training iterations $T$
\STATE Initialize planner policy $\pi_{\theta} \leftarrow \pi_{\theta}^{\text{off}}$
\FOR{$t = 1$ to $T$}
    \STATE $\mathcal{P}_{\text{on}} \leftarrow \emptyset$
    \FOR{each $(x_{\text{on}}, a_{\text{gold}}) \in \mathcal{D}_{\text{on}}$ (or a minibatch)}
        \STATE $x_{\text{on}} = (q, D, a_0, \hat{c})$
        \STATE Sample or decode a set of candidate plans $\mathcal{Y}$ from current policy $\pi_{\theta}(\cdot | x_{\text{on}})$
        \FOR{each $y_j \in \mathcal{Y}$}
            \STATE $\hat{a}_j \leftarrow \text{Executor}(x_{\text{on}}, y_j)$
            \STATE $s_j \leftarrow R(x_{\text{on}}, y_j)$
        \ENDFOR
        \STATE Induce preferences over $\mathcal{Y}$ using scores $\mathcal{S}$ and construct one or more triples $(x_{\text{on}}, y^{+}, y^{-})$
        \STATE Add all resulting triples to $\mathcal{P}_{\text{on}}$
    \ENDFOR
    \STATE Update $\pi_{\theta}$ on $\mathcal{P}_{\text{on}}$ using the DPO loss (Eq.~\eqref{eq:dpo})
\ENDFOR
\STATE \textbf{return} $\pi_{\theta}$
\end{algorithmic}
\end{algorithm}

\section{Details of Experimental Settings}
\label{app:details}

All experiments were conducted in a Python environment based on PyTorch. Model training and inference were implemented using the HuggingFace ecosystem, with distributed training and preference optimization supported by \texttt{accelerate}, \texttt{deepspeed}, \texttt{trl}, and parameter-efficient fine-tuning via \texttt{peft}~\cite{DBLP:conf/emnlp/WolfDSCDMCRLFDS20, DBLP:conf/kdd/RasleyRRH20}. Fast inference was enabled by \texttt{vllm}~\cite{kwon2023vllm}. Retrieval components relied on FAISS and dense embedding toolkits, with sparse retrieval baselines supported by \texttt{pyserini} and BM25 utilities~\cite{DBLP:conf/sigir/LinMLYPN21,DBLP:journals/corr/abs-2401-08281}. 

\section{DPO Optimization and Configuration.}
\label{app:dpo}

We trained the DPO objective using DeepSpeed ZeRO Stage~3 to enable memory-efficient optimization. Batch-related parameters, including the global training batch size, per-GPU micro-batch size, and gradient accumulation steps, were set to \texttt{auto} to allow DeepSpeed to adaptively determine optimal values based on available hardware resources. Parameter persistence thresholds were set to retain frequently accessed parameters in GPU memory when feasible, whereas 16-bit weights were collected only at model save time to minimize runtime overhead. Logging and diagnostic options were kept lightweight, with periodic step-level reporting and wall-clock breakdown disabled.

DPO training was performed using a single-stage schedule with a preference scaling coefficient $\beta=0.1$. Models were trained for one epoch using the AdamW optimizer with a learning rate of $5\times10^{-6}$ and a linear warmup over the first 10\% of training steps. We used a per-device batch size of 1 with gradient accumulation over two steps, yielding a small effective batch size consistent with prior DPO setups. Input sequences were truncated to a maximum length of 4096 tokens, with prompts capped at 2048 tokens. Mixed-precision training was enabled using bfloat16 when hardware support was available, and attention kernels were selected adaptively, with optional support for FlashAttention when available.

\section{Case Analysis}
\label{app:cases}
Figure~\ref{fig:error_vs_plan} illustrates a representative case highlighting how error categorization influences planning in agentic RAG. Given a factual question about banking regulation in India, the retriever surfaces passages mentioning several related statutes, including the \emph{Banking Regulation Act, 1949}, the \emph{Companies Act, 2013}, and the \emph{Reserve Bank of India Act, 1934}. The initial RAG response fails to directly identify the governing act, resulting in an erroneous answer. Under the golden error categorization, this failure is correctly attributed to an answer-level error rather than a retrieval deficiency, leading to a direct correction via answer regeneration. In contrast, an incorrectly predicted error categorization attributes the failure to insufficient or irrelevant retrieval, triggering unnecessary query rewriting and additional retrieval steps before regenerating the answer. RePAIR avoids such redundant actions by bypassing fine-grained misdiagnosis and directly selecting the appropriate action, demonstrating how accurate or taxonomy-free error handling can yield more efficient and effective plans.

\section{Prompt Templates}
\label{app:prompts}

We adopt a system prompt and a user prompt to guide the optimization of the RAG process. Both prompts are adapted from those used in RAG-Critic and are modified to better align with our settings. Specifically, the user prompt provides the RAG state, including the question, retrieved documents, prior model response, and error signal, and instructs the agent to generate only the minimal sequence of function calls necessary to resolve the error. The prompts emphasize concise, action-oriented planning without reliance on explicit error taxonomies.

\begin{lstlisting}[caption={System prompt for RAG optimization agent.},label={lst:system_prompt}]
You are an agent tasked with optimizing a Retrieval-Augmented Generation process. The goal is to improve the model's predictions by addressing issues flagged in the error_type. You are given the results from an initial RAG process, including a query, a list of retrieved documents, a prediction, and the identified error type. Your task is to optimize the current RAG process by selecting the appropriate functions and generating the corresponding Python code to fix the problem.

Available Functions

1. Retrieval(query: str, topk: int) -> List[str]
   Purpose: Retrieves the top-k most relevant documents for a given query.
   Parameters:
     - query (str): input query
     - topk (int): number of documents
   Returns:
     - list of documents sorted by relevance

2. RewriteQuery(query: str, instruction: str) -> List[str]
   Purpose: Rewrite the query to better match relevant documents.
   Instructions:
     - "clarify": make the query more specific
     - "expand": add context or related terms

3. DecomposeQuery(query: str) -> List[str]
   Purpose: Decompose the query into more specific sub-queries.

4. RefineDoc(query: str, doc: str, instruction: str) -> str
   Purpose: Refine a document when it is not directly relevant.
   Instructions:
     - "explain"
     - "summarize"

5. GenerateAnswer(query: str, docs: List[str],
                  additional_instruction: str = None) -> str
   Purpose: Generate the final answer using the selected documents.

You can directly use the provided variables as inputs to the functions. You may freely combine functions to improve performance.
\end{lstlisting}

\begin{lstlisting}[caption={User prompt for RAG optimization.},label={lst:user_prompt}]
Given the following information:

question = "{question}"
doc_list = {doc_list}
previous_pred = "{previous_pred}"

Error type of previous prediction:
{error_type}

Please carefully read the provided question, document list, previous answer, and the error type given by a teacher model. Your task is to generate Python code that calls the relevant functions to optimize the current RAG process and resolve the identified error.

The generated code should:
- Contain only function calls (no implementations)
- Use a minimal and necessary sequence of function executions
- End with: final_answer = GenerateAnswer(...)

Only output the code. Do not provide explanations.
\end{lstlisting}

\end{document}

%% file: latex/tables/table-dataset.tex
\begin{table}
\centering
\small
\begin{tabular}{llrr}
\toprule
\textbf{Dataset} & \textbf{Task} & \textbf{Train} & \textbf{Test} \\
\midrule
NQ    & Single-hop QA        & 79.1k & 8.7k \\
2Wiki  & Multi-hop QA         & 15.0k & 12.5k \\
WoW& Dialogue Generation & 63.7k & 3.0k \\
\bottomrule
\end{tabular}
\vspace{0.4em}
\caption{Benchmark dataset specifics.}
\label{tab:dataset_sizes}
\end{table}

%% file: latex/figure-example.tex
\begin{figure}[t]
\centering
\scriptsize

\mdfdefinestyle{ex}{
innerleftmargin=2pt, 
innerrightmargin=2pt,
innertopmargin=2pt,
innerbottommargin=2pt
}

\begin{mdframed}[style=ex]
\textbf{RAG Input State}\\
\textbf{Question:}\\
Which act governs the working of banking companies in India?\\[0.3em]
\textbf{Retrieved Passages:}\\
\emph{Banking Regulation Act, 1949;} \emph{Companies Act 2013;}
\emph{Reserve Bank of India Act, 1934}...\\[0.3em]
\textbf{Initial RAG Response:}\\
To determine which act governs...
\end{mdframed}
\vspace{-1em}
\begin{mdframed}[style=ex]
{\setlength\fboxsep{0.5pt}%
 \colorbox{DarkGreen}{\textcolor{black}{\textbf{Golden label error categorization}}}}\\
\textbf{High-level:}\\
Incomplete/Missing Response;\
Inaccurate/Misunderstood Response;\
Irrelevant/Off-Topic Response;\
Erroneous Information\\
\textbf{Fine-grained:}\\
Content--Context Misalignment;\
Entity/Concept Confusion;\
Specificity/Precision Errors;\
Erroneous Retrieval\\
\textbf{Planned Correction:}\\
{\setlength\fboxsep{0.5pt}%
 \colorbox{DarkGreen}{\textcolor{black}{\textsc{GenerateAnswer}}}}
\end{mdframed}
\vspace{-1em}
\begin{mdframed}[style=ex]
{\setlength\fboxsep{0.5pt}%
 \colorbox{DarkRed}{\textcolor{black}{\textbf{Incorrectly predicted error categorization}}}}\\
\textbf{High-level:}\\
Incomplete Information;\
Irrelevant Information\\
\textbf{Fine-grained:}\\
Insufficient or Incomplete Information Retrieval;\
Irrelevant Information Retrieval\\
\textbf{Planned Correction:}\\
{\setlength\fboxsep{0.5pt}%
 \colorbox{DarkRed}{\textcolor{black}{\textsc{RewriteQuery}}}},
{\setlength\fboxsep{0.5pt}%
 \colorbox{DarkRed}{\textcolor{black}{\textsc{Retrieval}}}},
{\setlength\fboxsep{0.5pt}%
 \colorbox{DarkGreen}{\textcolor{black}{\textsc{GenerateAnswer}}}}
\end{mdframed}
\vspace{-1em}
\begin{mdframed}[style=ex]
\textbf{RePAIR Planned Correction:}\\
{\setlength\fboxsep{0.5pt}%
 \colorbox{DarkGreen}{\textcolor{black}{\textsc{GenerateAnswer}}}}
\end{mdframed}
\caption{An example of correct vs.\ incorrect error categorization. Misclassification results in ineffective refinement.}
\label{fig:error_vs_plan}
\end{figure}